\documentclass{Interspeech2024}
\usepackage{textcomp}
\usepackage{amsmath}
\usepackage{tabularx} % 引入tabularx包
\usepackage{booktabs} % 如果您使用了booktabs包的命令，也需要引入
\usepackage{tabularray}
\usepackage{rotating} % 在导言区添加这个宏包
\usepackage{authblk}
\title{Unveiling the Potential of LLM-Based ASR on Chinese Open-Source Datasets}

\name[]{Xuelong}{Geng}
\name[]{Tianyi}{Xu}
\name[]{Kun}{Wei}
\name[]{Bingshen}{Mu}
\name[]{Hongfei}{Xue}
\name[]{He}{Wang}
\name[]{Yangze}{Li}
\name[]{Pengcheng}{Guo}
\name[]{Yuhang}{Dai}
\name[]{Longhao}{Li}
\name[]{Mingchen}{Shao}
\name[]{Lei}{Xie}

\address{
  Audio, Speech and Language Processing Group {(ASLP@NPU)},  \\ Northwestern Polytechnical University, Xi'an
}

\email{xlgeng@mail.nwpu.edu.cn; lxie@nwpu.edu.cn}


\begin{document}
\maketitle 

\begin{abstract}
    
    % todo: These research findings not only showcase the immense developmental potential of the ASR system combined with LLM but also pave new avenues for future research in speech recognition technology.这一句用法可能有些夸张，另外not only but also这样的用法也并不简洁
    %a training methodology, termed ‘three-stage training 有些重复，建议写为 we introduce a three stage training 
    % The culmination of our research is a model, meticulously assembled 太过文艺 , unparalleled  无可匹敌，太过夸张 建议写为
    % 这不是英语作文，不会因为你用了“高难词汇”加分，简单准确表达出意思为上
% current best performance 可写为SOTA
%academic community 写为 community即可，有些重复
 % remain nascent 过于文艺，Investigations into LLM-based ASR systems using extensive open-source Chinese datasets remain nascent.  替换为 Few 
Large Language Models (LLMs) have demonstrated unparalleled effectiveness in various NLP tasks, and integrating LLMs with automatic speech recognition (ASR) is becoming a mainstream paradigm. Building upon this momentum, our research delves into an in-depth examination of this paradigm on a large open-source Chinese dataset. Specifically, our research aims to evaluate the impact of various configurations of speech encoders, LLMs, and projector modules in the context of the speech foundation encoder-LLM ASR paradigm. Furthermore, we introduce a three-stage training approach, expressly developed to enhance the model's ability to align auditory and textual information. The implementation of this approach, alongside the strategic integration of ASR components, enabled us to achieve the SOTA performance on the AISHELL-1, Test\_Net, and Test\_Meeting test sets. Our analysis presents an empirical foundation for future research in LLM-based ASR systems and offers insights into optimizing performance using Chinese datasets. We will publicly release all scripts used for data preparation, training, inference, and scoring, as well as pre-trained models and training logs to promote reproducible research.

\end{abstract}

\section{Introduction}
Large Language Models (LLMs)~\cite{llama,bert} have emerged as a formidable force in artificial intelligence, showcasing unparalleled proficiency in understanding and generating human language. Drawing from this strength, researchers have begun to merge the prowess of LLMs with various fields, including automatic speech recognition (ASR), where their integration has led to notable performance improvement~\cite{achiam2023gpt,team2023gemini,speechgpt}. Specifically, ASR, a task that intricately intertwines acoustic modeling with language modeling, has conventionally employed language models like n-grams~\cite{ngram-1,ngram-2,shallow-fusion} or neural network language models (NNLMs)~\cite{nnlm-1,nnlm-2,deep-fusion,component-fusion}.  However, the advent of LLMs offers a compelling alternative to the language component of ASR, drawing from their superior ability to understand and predict linguistic patterns by scaling up data and parameters.

Research efforts to integrate LLMs with ASR systems generally fall into two categories. The first strategy involves connecting LLMs with pre-trained ASR models, wherein the ASR-generated text is directly fed to the LLM to serve as a prompt for downstream tasks~\cite{linkpara1_audio_gpt,linkpara-2,linkpara-4} or to facilitate error correction~\cite{linkpara-5}. However, this coarse-grained integration may result in a substantial loss of acoustic information. It may suffer compounding errors from the initial ASR stage, potentially leading to an exacerbation of inaccuracies. In contrast, the second approach adopts audio-text cross-modal LLMs, which embrace the auditory modality by employing an encoder network to process the speech and generate embeddings that are subsequently provided to a decoder-only LLM~\cite{speechgpt,decoder-only-asr,listen_think_anderstand,salmonn,qwen-audio, simple-asr-llm}. This framework strives for a tighter coupling between acoustic cues and linguistic context, aiming to yield a better interpretation of speech. 
Through a series of studies, the paradigm of augmenting a speech foundation model with an LLM through projector modules has emerged as the prevailing framework in the current LLM-based speech recognition research. Specifically, SALMONN~\cite{salmonn} applies Whisper~\cite{whisper} to extract semantic content and BEATs~\cite{beats} for audio event information, culminating in a robust perception of human speech, music, and audio events. Qwen-Audio~\cite{qwen-audio} implements Whisper as the exclusive encoder, utilizing structured task directives to enhance the model's performance across various audio tasks. SLAM-ASR~\cite{simple-asr-llm} leverages a linear layer as the projector module and achieves SOTA performance on the English 960-hour LibriSpeech~\cite{libispeech} task.

Following the inspiring results of these studies, we aim to investigate further the potential of the speech foundation encoder plus LLM decoder paradigm on a large-scale open-source Chinese dataset. Specifically, with over 11,000 hours of Chinese speech data from various corpora, we examine the impact of different projectors, speech encoders, and LLMs on Mandarin ASR performance within this paradigm.  Concurrently, we utilize a three-stage training approach designed to enhance the learning of the alignment between auditory and textual modalities. From experiments, we draw the following major conclusions: (1) For the speech encoder, Whisper~\cite{whisper} is more robust  but have lower plasticity compared to HuBERT~\cite{hubert}.  (2) For the projector, the Transformer’s learning ability is better than the Qformer~\cite{blip} in the speech recognition task.  (3) For the LLM,  the performance of the LLM-integrated ASR system is positively correlated to the LLM's proficiency in that specific language -- Mandarin here.  (4) Our three-stage training approach can effectively align the pre-trained acoustic modeling capability of the speech foundation model with the language modeling capability of LLMs, using a relatively smaller Chinese dataset and achieve SOTA performance on the AISHELL-1, Test\_Net, and Test\_Meeting test datasets.

 We will provide reproducible recipes encompassing the entire pipeline, including data preparation, training, inference, and scoring. Furthermore, we will release pre-trained models, enabling researchers to delve into the specifics of the training process and gain valuable insights for their own investigations\footnote{https://github.com/gengxuelong/wenet\_LLM\_from\_ASLP}.

\section{Method}

\begin{figure}[h]
  \centering
  \includegraphics[width=\linewidth,height=4cm]{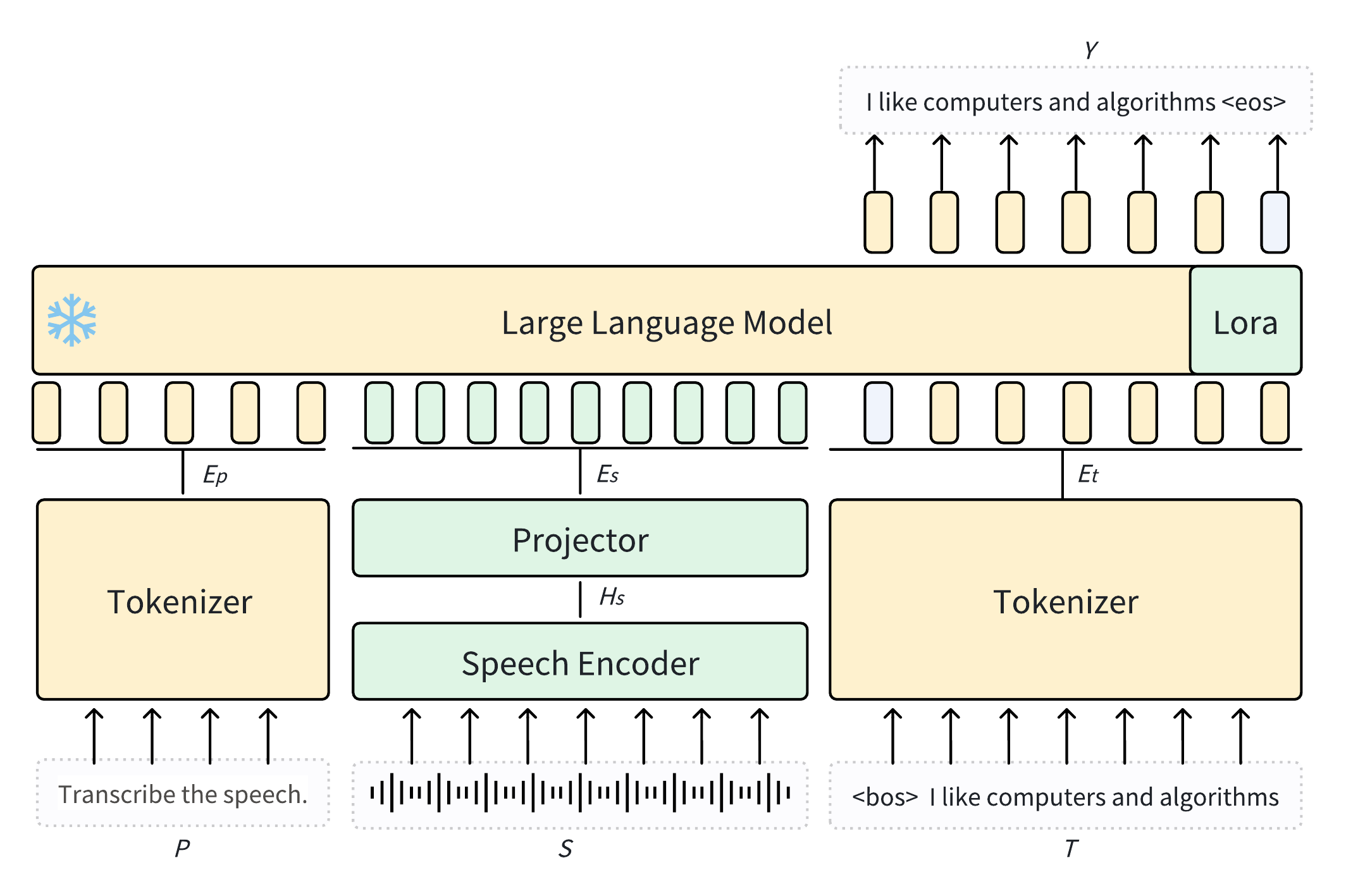}
  \caption{Overall model structure. The embedding sequence generated by the projector is concatenated with the text embedding sequence. This concatenated sequence is fed directly into the decoder-only LLM, predicting the next token.}
  \label{fig:0001}
\end{figure}

% \subsection{Overall Architecture}
\label{sec:2.1}
As shown in Figure~\ref{fig:0001}, the architecture for investigation is simply a speech encoder with an LLM. For each sample, the given prompt for transcribing (i.e., transcribe the speech), the speech utterance, and the corresponding transcript during training are denoted as \( P \), \( S \) and \( T \), respectively.  We tokenize the prompt and the transcript using the tokenizer and embedding matrix of the LLM to obtain feature vector sequences \( E_p \) and \( E_t \) as:
\begin{align}
E_p &= \text{Embedding}(\text{Tokenizer}(P)),\label{eq:multi1} \\
E_t &= \text{Embedding}(\text{Tokenizer}(T)).\label{eq:multi2}
\end{align}
For the input audio \( S \), we first extract features by passing the audio through a speech encoder to obtain encoder output \( H_s \), denoted as:
\begin{equation}
H_s = \text{Encoder}(S).
\label{myequation1}
\end{equation}
Then, \( H_s \) is passed to a projector and further goes through a linear layer to obtain a feature sequence \( E_{\text{s}} \) with the same dimensionality as the input to the LLM, denoted as:
\begin{equation}
E_{\text{s}} = \text{Linear}(\text{Projector}(H_s)),
\label{myequation2}
\end{equation}
where the dimension of the feature output by the projector is the same as that of the speech encoder, and the linear layer is responsible for mapping the feature dimension to the embedding dimension of the LLM.
Next, we concatenate \( E_{\text{s}} \),  \( E_p \), and \( E_t \) to obtain the final feature and pass it to the LLM to obtain the output transcript \(Y\), denoted as:
\begin{equation}
Y = \text{LLM}(\text{Regulation}(E_p, E_{\text{s}}, E_t)).
\label{eq:1}
\end{equation}

% \( E_a \) passed to the LLM, denoted as:
% \begin{equation}
% E_a = \text{Regulation}(E_p, E_{\text{sl}}, E_t).
% \label{myequation3}
% \end{equation}
% Finally, we pass \( E_a \) to the LLM to obtain the output transcript \(Y\), denoted as:
% \begin{equation}
% Y = \text{LLM}(E_a).
% \label{eq:1}
% \end{equation}

\section{Experimental Setup}
\subsection{Datasets}
\textbf{Training Set:} We use over 11,000 hours of Chinese data from four corpora, including WenetSpeech~\cite{wenetspeech}, AISHELL-1~\cite{aishell1},  AISHELL-2~\cite{aishell2},  and AISHELL-4~\cite{aishell4}, as shown in Table~\ref{tab:02}. It is worth noting that, we use data from WenetSpeech with corrected transcripts\footnote{https://github.com/wenet-e2e/WenetSpeech/discussions/54}. To balance the data ratio between the AISHELL datasets and the WenetSpeech dataset, we replicated the AISHELL datasets three times during training. 

\noindent \textbf{Testing Set:} To reduce bias and identify unique challenges, we test on nine open-source and two internal test sets, as shown in  Table~\ref{tab:02}. Specifically, the open test sets include AISHELL-1~\cite{aishell1}, AISHELL-2~\cite{aishell2}, Test\textunderscore{}Net~\cite{wenetspeech}, Test\textunderscore{}Meeting~\cite{wenetspeech}, SPEECHIO\_0, SPEECHIO\_1, SPEECHIO\_2, SPEECHIO\_3, SPEECHIO\_4\footnote{SPEECHIO\_\textsuperscript{*} means SPEECHIO\_ASR\_ZH0000\textsuperscript{*}}~\cite{speechio}. The internal noisy and accent test sets have different acoustic conditions (noisy and accented speech) from the open test sets.

\begin{table}[h]
\centering
\caption{The training set and testing set used in this study.}
\label{tab:02}
\scalebox{0.9}{
\begin{tblr}{
width = 0.37\textwidth,
colspec = {Q[m,c,0.02\textwidth]Q[m,c,0.15\textwidth] Q[m,c,0.03\textwidth] Q[m,c,0.17\textwidth]},
  cell{2}{1} = {r=4}{},
  cell{6}{1} = {r=9}{},
  hline{1,17} = {-}{0.08em},
  hline{2,6} = {-}{},
}
      & Dataset          & Hours  & Scenario             \\
Train & WenetSpeech      & 10,000 & Multiple domains     \\
      & AISHELL-1(train) & 178    &  Read speech    \\
      & AISHELL-2(train) & 1,000  &  Read speech      \\
      & AISHELL-4(train) & 120    & Conference           \\
Test  & AISHELL-1(test)  &  5      & Read speech    \\
      & AISHELL-2(test)  &   15     &Read speech      \\
      & Test\_Net        &   23     & Multiple domains     \\
      & Test\_Meeting    &  15      & Multiple domains     \\
      & SPEECHIO\_0      &    1    & Conference recording\\
      & SPEECHIO\_1      &     9   &  Evening TV news                \\
      & SPEECHIO\_2      &      3  & Financial  news        \\
      & SPEECHIO\_3      &       2.7 & Football commentary       \\
      & SPEECHIO\_4      &      2.7  & Public Speech \\
      & Internal Noisy & 0.5 & Child speech w/ noise\\
      &Internal Accents &1 & Strong accents\\
\end{tblr}
}
\end{table}

\subsection{Components}
\subsubsection{LLM}
As shown in Table~\ref{tab:llmscomparison}, we experimented with two LLMs: Atom-7B\footnote{https://huggingface.co/FlagAlpha/Atom-7B}, and Baichuan2-7B-Chat\footnote{https://huggingface.co/baichuan-inc/Baichuan2-7B-Chat}. Specifically, Atom-7B represents an LLM fine-tuned on Llama2-7B~\cite{llama}, with over 1T Chinese characters of data. According to the Chinese LLM Benchmark (CiLB)\footnote{https://github.com/jeinlee1991/chinese-LLM-benchmark?tab=readme-ov-file}, Baichuan2-7B-Chat has the best overall performance among all 7B models for the Chinese domain. Concurrently, the SLAM-ASR~\cite{simple-asr-llm} suggests that the chat models are generally superior to their ordinary counterparts. In light of these considerations, we select Baichuan2-7B-Chat which  is trained with 2.6T tokens as another LLM for our study.
% Following the results in ~\cite{simple-asr-llm}, which suggests that the chat models are generally superior to their ordinary counterparts, we also investigate Baichuan2-7B-Chat. Baichuan2-7B-Chat is trained with 2.6T tokens and has the best overall performance among all 7B models according to the Chinese LLM Benchmark (CiLB)\footnote{https://github.com/jeinlee1991/chinese-LLM-benchmark?tab=readme-ov-file}.

% \usepackage{color}
% \usepackage{tabularray}
\begin{table}[htbp]
\centering
\caption{Details of two 7B LLMs in this study.}
\label{tab:llmscomparison}
\scalebox{0.9}{
\begin{tblr}{
width = 0.37\textwidth,
colspec = {Q[m,c,0.07\textwidth] Q[m,c,0.18\textwidth] Q[m,c,0.14\textwidth]},
  hline{1,4} = {-}{0.08em},
  hline{2} = {-}{}
}
LLM  & \textbf{Atom-7B (Llama2 based)} & \textbf{ Baichuan2-7B-chat}             \\
Vocabulary Size    & 65,000  & 125,696  \\
Training Data& over 1T characters of Chinese data drawn from diverse domains & a high-quality set with 2.6T Chinese characters \\
\end{tblr}
}
\end{table}

\subsubsection{Speech foundation encoder}
We investigate encoders from two speech foundation models: Whisper~\cite{whisper} and HuBERT~\cite{hubert}, representing supervised and self-supervised models, respectively. Whisper~\cite{whisper} is a supervised speech foundation model trained with 680,000 hours of labeled multilingual data. We use the large-v2 version of Whisper with a model parameter count of 640M and a feature dimension of 1280. This Whisper model is fine-tuned with the training set described in Section 3.1 for 3 epochs. HuBERT~\cite{hubert} is an unsupervised speech foundation model that learns speech representations from unlabeled data.  Here, we use an open-source HuBERT-large model trained with WenetSpeech\footnote{https://github.com/TencentGameMate/chinese\_speech\_pretrain}, with 317M parameters and 1024 feature dimensions.

% We add a linear layer map of the feature dimension from 1024 to 1280. 

%The speech features are passed through the multi-layer self-attention module to learn the alignment information between audio modality and text modality.

\subsubsection{Projector}
We investigate two types of projectors, Qformer~\cite{blip} and Transformer~\cite{transformer}. For Qformer, we follow the configuration in SALMONN~\cite{salmonn}, and set the window length to 1, the number of trainable queries to 1, and the number of layers to 2. The total number of parameters of Qformer is 51M. We use 4 self-attention layers of the standard Transformer layers for the Transformer projector, and the total number of parameters is also about 51M. Both Qformer and Transformer have the same feature dimensions as the Whisper, with a linear mapping layer behind them to align the feature dimensions with the LLM. When HuBERT is employed as the encoder, a linear layer is utilized to map the feature dimension from 1024 to 1280.

\subsection{Training strategy}
\label{sec:3.1}
We introduce a three-stage training strategy to enhance the model’s capacity to align auditory and textual information. Initially, we attempted to simultaneously unfreeze the speech encoder, projector, and LLM LoRA~\cite{lora} matrix during training. However, the model failed to converge and had a poor performance, as shown later in Table~\ref{tab:table2}. We hypothesize that this is because our dataset was relatively small compared to the data used to train the encoder and the LLM,  which resulted in a mismatch between the representations of speech and text within LLMs, similar to the results in ~\cite{simple-asr-llm}. Thus, we utilize a three-stage training approach to alleviate this problem: First, we train only the projector, freezing all other components, aiming to train the projector to align between auditory and textual modalities.  Second, we unfreeze the encoder, focusing solely on the encoder. This step aims to adapt the speech encoder to our datasets. Third, we fine-tune the LLM with LoRA (Low-Rank Adaptation)~\cite{lora}. This step aims to adapt the LLM output to the style of ASR transcription.

\subsection{Implementation details}
\label{sec:3.2}
%During training, our input format is as follows: ``\textless P\textgreater:\textless Speech\textgreater\textless S\textgreater\textless /Speech\textgreater \textbackslash nASSISTANT:\textless O\textgreater'', where \textless P\textgreater ~stands for prompt and always ``transcribe the speech", \textless S\textgreater ~represents the speech features processed through the connecting layer, and \textless O\textgreater ~denotes the corresponding label information for the speech. Specific details for training can be found in Section~\ref{sec:2.1}.

We train our model using eight A6000 GPUs.
We use the AdamW optimizer~\cite{adamw} with the following hyperparameters: lr = 5.0e-05, beta = (0.9, 0.99), eps = 1.0e-06, and weight\textunderscore{}decay = 0.01. Regarding the learning rate, we employ the warmup scheduler~\cite{transformer}, which implements inverse square root decay. In our experiments, we set the warmup steps to 2000. To mitigate potential issues related to gradient explosion during training, we apply gradient clipping~\cite{grad_clip} with a threshold value set to 5. This ensures that gradients over 5 or below -5 are clipped to 5 or -5, respectively. Additionally, we adopt a gradient accumulation of 14 to increase the effective batch size without a proportional increase in memory usage. Furthermore, we employ a dynamic batch type where the number of utterances in each mini-batch is determined based on the total number of frames or sample points rather than a fixed value. This parameter is set to 400,000, representing the sample points per batch.
Regarding the training approach, when training the LLM, we freeze the LLM body and only update the LLM using LoRA fine-tuning~\cite{lora}. We configure LoRA with alpha = 32, rank = 8. The alpha parameter controls the weight of the LoRA matrix, and the rank parameter determines the dimension of the low-rank matrix.  All experiments in this work follow these configurations unless otherwise specified.

\section{Experimental Results}

\subsection{Experiments on projector}
We compare the effects of two types of projectors, Qformer and Transformer. In our experiment, we fix the encoder as HuBERT and LLM as Atom-7B and only unfreeze the projector module for one epoch training to compare the different effects of Qformer and Transformer, respectively. Although we expect Qformer to perform better than transformer similar to Blip \cite{blip} and SALMONN~\cite{salmonn}, results in Table~\ref{tab:table2} show otherwise. We hypothesize this is because Qformer was originally designed to accommodate the unique data structures in image processing, and should be redesigned to better adapt to speech for future research. 

\begin{table}[h]
\centering
\caption{Comparison of projector modules in terms of CER\%($\downarrow$). The encoder is fixed to HuBERT and the LLM is fixed to Atom-7B. Only one epoch is trained. Here, `all' represents the simultaneous unfreezing of the speech encoder, projector, and LLM LoRA matrix during training. }
\label{tab:table2}
\scalebox{0.9}{
\begin{tblr}{
width = 0.37\textwidth,
colspec = {Q[m,c,0.16\textwidth]Q[m,c,0.06\textwidth] Q[-0.01\textwidth]Q[m,c,0.05\textwidth] Q[m,c,0.07\textwidth]},
% row{even} = {c},
  cell{1}{1} = {r=2}{},
  cell{1}{2} = {c=3}{},
  cell{2}{4} = {c=2}{},
  hline{1,13} = {-}{0.08em},
  hline{2} = {2-5}{},
  hline{3} = {4-5}{},
  hline{3} = {2-2}{},
  hline{4} = {1-5}{},
}
Dataset         & Projector &                               &              \\
                & \textbf{\textbf{Qformer}} && \textbf{\textbf{Transformer}} &              \\
         unfrozen layers       & projector             && projector                 & all  \\
AISHELL-1(test) & 4.63                     & & \textbf{3.24}                          & 8.77         \\
AISHELL-2(test) & 6.92                     & & \textbf{5.87}  & 10.25        \\
Test\_Net       & 15.39                    & & \textbf{11.93 }                        & 23.35        \\
Test\_Meeting   & 22.50                    & & \textbf{17.14}                         & 30.83        \\
SPEECHIO\_0     & 12.35                    & & \textbf{10.14}                         & 12.96        \\
SPEECHIO\_1     &\textbf{ 4.32 }                    & & 4.79                          & 11.21        \\
SPEECHIO\_2     & 11.17                    & & \textbf{8.44}                          & 13.21        \\
SPEECHIO\_3     & 8.23                      && \textbf{7.16  }                        & 12.55        \\
SPEECHIO\_4     & 7.74                      &&\textbf{ 7.35  }                        & 10.89        
\end{tblr}
}
\end{table}

\subsection{Experiments on speech encoder}
Next, we compare the two speech encoders, Whisper and HuBERT. We fix the projector to Transformer, the LLM to Atom-7B, and then train one epoch on stage 1 and two epochs on stage 2.  According to results in Table~\ref{tab:0002}, HuBERT outperforms Whisper in the open-source test sets, and Whisper outperforms HuBERT in the internal noisy and accent test sets. This suggests that Whisper is more robust yet has low plasticity and is harder to adapt to new domains compared to HuBERT, presumably due to its significantly larger dataset in the original training and the large parameter size.

\begin{table}[h]
\centering
\caption{Comparison of speech encoders in terms of CER\%($\downarrow$). The projector is fixed to the Transformer and the LLM is fixed to Atom-7B, training three epochs.}
\label{tab:0002}
\scalebox{0.9}{
\begin{tblr}{
width = 0.39\textwidth,
colspec = {Q[m,c,0.16\textwidth] Q[m,c,0.12\textwidth] Q[m,c,0.12\textwidth]},
  cell{1}{1} = {r=2}{},
  cell{1}{2} = {c=2}{},
  hline{1,14} = {-}{0.08em},
  hline{2} = {2-3}{0.03em},
  hline{3} = {1}{0.03em},
  hline{3} = {2-3}{},
}
Dataset             & Speech Encoder &                           \\
                    & \textbf{\textbf{Whisper}} & \textbf{\textbf{HuBERT}} \\
AISHELL-1(test)           & 4.15                      & \textbf{3.17  }                    \\
AISHELL-2(test)         & 5.72                      & \textbf{5.59  }                    \\
Test\_Net           & 16.30                     & \textbf{11.78 }          \\
Test\_Meeting       & 16.53                     & \textbf{16.00 }                    \\
SPEECHIO\_0         &\textbf{ 7.58                      }& 8.33                      \\
SPEECHIO\_1         &\textbf{ 4.10}                      & 4.43                      \\
SPEECHIO\_2         & 8.29                      & \textbf{7.15       }               \\
SPEECHIO\_3         & 7.04                      & \textbf{6.51     }                 \\
SPEECHIO\_4         & 6.87                      & \textbf{6.33     }                 \\
Internal Noisy &\textbf{ 38.41   }                  & 65.72                     \\
Internal Accents & \textbf{33.10  }                   & 49.78                     
\end{tblr}
}
\end{table}

\subsection{Experiments on LLM}

Finally, we compare different LLMs, Atom-7B and Baichuan2-7B-Chat. We fix the speech encoder as HuBERT and the projector as Transformer and only unfreeze the projector for training with one epoch. To activate the chat capability of LLM when using the Baichuan2-7B-Chat model, The chat-related prompt of the Baichuan2-7B-Chat model is used for training and decoding.

We find that the ASR performance positively correlates to the LLM's performance in Chinese NLP tasks. Specifically, the Baichuan2-7B-Chat model, which has a better rating according to the CiLB has the lower CER, as shown in Table~\ref{tab:0003}. This suggests that the performance of a speech foundation encoder-LLM ASR framework is positively correlated to the LLM's proficiency in that specific language. 
\begin{table}
\centering
\caption{Comparison of LLMs in terms of CER\%($\downarrow$). The projector is fixed to Transformer and the encoder is fixed to HuBERT. Only one epoch of the projector is trained.}
\label{tab:0003}
\scalebox{0.9}{
\begin{tblr}{
  cells = {c},
  cell{1}{1} = {r=2}{},
  cell{1}{2} = {c=2}{},
  hline{1,12} = {-}{0.08em},
  hline{1,3,12} = {-}{},
  hline{2} = {2-3}{},
}
Dataset         & LLM                       &                                     \\
                & \textbf{\textbf{Atom-7B}} & \textbf{\textbf{Baichuan2-7B-chat}} \\
AISHELL-1(test) & \textbf{3.24}                     &  4.28                     \\
AISHELL-2(test) & 5.87                      & \textbf{5.51 }                      \\
Test\_Net       & 11.93                     & \textbf{ 10.22 }                    \\
Test\_Meeting   & 17.14                    & \textbf{11.51 }                     \\
SPEECHIO\_0     & 10.14                     & \textbf{8.27 }                      \\
SPEECHIO\_1     & 4.79                      & \textbf{ 2.34 }                     \\
SPEECHIO\_2     & 8.44                     & \textbf{6.48 }                      \\
SPEECHIO\_3     & 7.16                      & \textbf{2.86}                       \\
SPEECHIO\_4     & 7.35                      & \textbf{4.43 }                      
\end{tblr}
}
\vspace{-10pt}
\end{table}

\subsection{Optimal configuration}
We compare our results to current SOTA models, Parafomrer-large~\footnote{https://huggingface.co/funasr/Paraformer-large}~\cite{paraformer} and Qwen-Audio~\footnote{https://huggingface.co/Qwen/Qwen-Audio}~\cite{qwen-audio}. We follow the three-stage training described in section~\ref{sec:3.1}, wherein the model undergoes one epoch of training in stage one, followed by two epochs in each subsequent stage. Through experiments in Table~\ref{tab:0004}, we find that the setup using HuBERT as encoder, Transformer as projector, and Baichuan2-7B-Chat as LLM performs the best, surpassing models that have much larger training data (Parafomrer-large 60,000 hours and Qwen-Audio 30,000 hours) on AISHELL-1, Test\_Net, and Test\_Meeting. 

We also compare our results to vanilla fine-tuned Whisper models and U2++ conformer model~\cite{u2++}. We use the same training set to fine-tune Whisper-large-v2, while the WeNet team\cite{wenet} fine-tunes Whisper-large-v3 using WenetSpeech (corrected transcript). Additionally, we train a standard U2++ conformer model using the same training data. We initialize the model with parameters from the WeNet team's publicly available model~\footnote{https://github.com/wenet-e2e/wenet/blob/main/docs/pretrained\_models.en.md}, which is trained solely on WenetSpeech data, and train it for 2 epochs using our training data.
The experimental results in Table \ref{tab:0004} show that our proposed model performs significantly better than the Whisper series models and U2++ conformer model. This indicates that the performance could not be achieved by the speech foundation model encoder alone, and the LLM helped to improve the performance.

\begin{table}[htbp]
\centering
%\textbf{Pa} stands for paraformer-large~\cite{paraformer}, \textbf{Qwen} stands for qwen-audio~\cite{qwen-audio}, \textbf{W1} represents the Whisper~\cite{Whisper} model fine-tuned with all parameters tuned using the large-v2 version, and \textbf{W2} represents the Whisper model fine-tuned with asll parameters tuned using the large-v3 version.
\caption{Comparison with popular models in terms of CER\%($\downarrow$). \textbf{Whisper-L-v2} and \textbf{Whisper-L-v3} have both undergone fine-tuning. Here, 'AISL1' refers to the AISHELL-1 test set, while 'AISL2' refers to the AISHELL-2 test set. 'Test\_N' and 'Test\_M' represent the Test\_Net and Test\_Meeting test sets, respectively. 'SIO\_' denotes the SPEECHIO\_ test sets. } 
\label{tab:0004}
\scalebox{0.9}{
\begin{tabular}{
>{\centering\arraybackslash}p{0.04\textwidth}
                >{\centering\arraybackslash}p{0.04\textwidth}
                >{\centering\arraybackslash}p{0.03\textwidth}
                >{\centering\arraybackslash}p{0.06\textwidth}
                >{\centering\arraybackslash}p{0.06\textwidth}
                >{\centering\arraybackslash}p{0.04\textwidth}
                >{\centering\arraybackslash}p{0.03\textwidth}
                }
\toprule
Dataset & \textbf{Para-former} & \textbf{Qwen-Audio} & \textbf{Whisper-L-v2} & \textbf{Whisper-L-v3} & \textbf{U2++} & \textbf{Ours} \\
\midrule

AISL1  &1.95   & 1.30& 3.18   & 5.02 &1.44    &\textbf{0.95}   \\ 
AISL2  &\textbf{3.01 }  & 3.18&  4.49  &  6.60    &3.92& 3.50   \\ 
Tset\_N & 6.74 & 9.50&  9.08  & 6.54& 8.32&  \textbf{6.06}  \\ 
Test\_M &6.97 & 10.87 &  9.73 & 8.96& 8.84&\textbf{6.26} \\ 
SIO\_0 &\textbf{2.55}& 5.08& 3.44   &  4.05&3.38     &3.05   \\ 
SIO\_1 &\textbf{0.49}& 1.17& 1.98   & 3.17  & 1.87 &1.58   \\ 
SIO\_2 & \textbf{3.23} & 5.68 &  5.27     & 6.25  & 4.90  &3.73   \\ 
SIO\_3 & \textbf{1.13} & 2.81& 4.86  & 7.09  & 3.88  &2.42   \\ 
SIO\_4 & \textbf{1.33} & 3.99 &3.63    & 4.29  & 3.59  &3.29   \\
\bottomrule
\end{tabular}
}
\end{table}

% SIO\_2 &1.95& \textbf{3.23} & 5.68 &  5.27     & 6.25   &3.73   \\ 
% SIO\_3 &1.95& \textbf{1.13} & 2.81& 4.86  & 7.09   &2.42   \\ 
% SIO\_4 &1.95& \textbf{1.33} & 3.99 &3.63    & 4.29   &3.29   \\

%problem: model proposed by us 换为了  our proposed model
%todo: In this article, we start by comparing different components of LLM-based ASR systems 换成了we compared, 要是有篇幅把start by ,then ... finally... 重新概括一下可以这么用，不然就单纯的总括一些我们干了什么

\section{Conclusion}
In this paper, we investigate various component configurations in speech-encoder LLM-decoder ASR systems using 11,000-hour open-source Mandarin data. For the speech encoder, Whisper is more robust but has lower plasticity compared to compared to HuBERT. For the projector, the Transformer's learning ability is better than the Qformer in speech recognition tasks. For the LLM,  the performance of the LLM integrated system is positively correlated to the LLM's proficiency in that specific language. We utilize a three-stage training method, which optimizes the alignment of auditory and textual modalities. Under the optimal combination (Hubert+Transformer+Baichuan-7B-Chat), our proposed model has obtained SOTA results on AISHELL-1, Test\_Net, and Test\_Meeting test datasets. By open-sourcing our recipes and pre-trained models, we hope our research can promote further exploration of LLM-based ASR research.

\bibliographystyle{IEEEtran}
\bibliography{mybib}

\begin{thebibliography}{10}
\providecommand{\url}[1]{#1}
\csname url@samestyle\endcsname
\providecommand{\newblock}{\relax}
\providecommand{\bibinfo}[2]{#2}
\providecommand{\BIBentrySTDinterwordspacing}{\spaceskip=0pt\relax}
\providecommand{\BIBentryALTinterwordstretchfactor}{4}
\providecommand{\BIBentryALTinterwordspacing}{\spaceskip=\fontdimen2\font plus
\BIBentryALTinterwordstretchfactor\fontdimen3\font minus \fontdimen4\font\relax}
\providecommand{\BIBforeignlanguage}[2]{{%
\expandafter\ifx\csname l@#1\endcsname\relax
\typeout{** WARNING: IEEEtran.bst: No hyphenation pattern has been}%
\typeout{** loaded for the language `#1'. Using the pattern for}%
\typeout{** the default language instead.}%
\else
\language=\csname l@#1\endcsname
\fi
#2}}
\providecommand{\BIBdecl}{\relax}
\BIBdecl

\bibitem{llama}
H.~Touvron, T.~Lavril, G.~Izacard, X.~Martinet, M.~Lachaux, T.~Lacroix, B.~Rozi{\`{e}}re, N.~Goyal, E.~Hambro, F.~Azhar, A.~Rodriguez, A.~Joulin, E.~Grave, and G.~Lample, ``{LLaMA: Open and Efficient Foundation Language Models},'' \emph{{CoRR}}, 2023.

\bibitem{bert}
J.~Devlin, M.~Chang, K.~Lee, and K.~Toutanova, ``{BERT: Pre-training of Deep Bidirectional Transformers for Language Understanding},'' in \emph{{NAACL-HLT}}, 2019.

\bibitem{achiam2023gpt}
OpenAI, ``{GPT-4 Technical Report},'' \emph{{CoRR}}, 2023.

\bibitem{team2023gemini}
G.~Team, R.~Anil, S.~Borgeaud, Y.~Wu, J.-B. Alayrac, J.~Yu, R.~Soricut, J.~Schalkwyk, A.~M. Dai, A.~Hauth \emph{et~al.}, ``{Gemini: a family of highly capable multimodal models},'' \emph{{CoRR}}, 2023.

\bibitem{speechgpt}
D.~Zhang, S.~Li, X.~Zhang, J.~Zhan, P.~Wang, Y.~Zhou, and X.~Qiu, ``{SpeechGPT: Empowering Large Language Models with Intrinsic Cross-Modal Conversational Abilities},'' in \emph{{EMNLP}}, 2023.

\bibitem{ngram-1}
R.~Ma, X.~Wu, J.~Qiu, Y.~Qin, H.~Xu, P.~Wu, and Z.~Ma, ``{Internal Language Model Estimation Based Adaptive Language Model Fusion for Domain Adaptation},'' in \emph{{ICASSP}}, 2023.

\bibitem{ngram-2}
M.~Jung, O.~Kwon, S.~Seo, and S.~Seo, ``{Blank Collapse: Compressing {CTC} emission for the faster decoding},'' \emph{{CoRR}}, 2022.

\bibitem{shallow-fusion}
D.~Zhao, T.~N. Sainath, D.~Rybach, P.~Rondon, D.~Bhatia, B.~Li, and R.~Pang, ``{Shallow-Fusion End-to-End Contextual Biasing},'' in \emph{{Interspeech }}, 2019.

\bibitem{nnlm-1}
Z.~Liu, K.~Li, S.~Bakshi, and F.~Peng, ``{Private Language Model Adaptation for Speech Recognition},'' \emph{{CoRR}}, 2021.

\bibitem{nnlm-2}
S.~Deena, M.~Hasan, M.~Doulaty, O.~Saz, and T.~Hain, ``{Recurrent Neural Network Language Model Adaptation for Multi-Genre Broadcast Speech Recognition and Alignment},'' \emph{{ACM}}, 2019.

\bibitem{deep-fusion}
K.~Irie, A.~Zeyer, R.~Schlüter, and H.~Ney, ``{Language Modeling with Deep Transformers},'' in \emph{Interspeech}, 2019.

\bibitem{component-fusion}
C.~Shan, C.~Weng, G.~Wang, D.~Su, M.~Luo, D.~Yu, and L.~Xie, ``{Component Fusion: Learning Replaceable Language Model Component for End-to-end Speech Recognition System},'' in \emph{{ICASSP}}, 2019.

\bibitem{linkpara1_audio_gpt}
R.~Huang, M.~Li, D.~Yang, J.~Shi, X.~Chang, Z.~Ye, Y.~Wu, Z.~Hong, J.~Huang, J.~Liu, Y.~Ren, Y.~Zou, Z.~Zhao, and S.~Watanabe, ``{AudioGPT: Understanding and Generating Speech, Music, Sound, and Talking Head},'' in \emph{{AAAI}}, 2024.

\bibitem{linkpara-2}
Y.~Shen, K.~Song, X.~Tan, D.~Li, W.~Lu, and Y.~Zhuang, ``{HuggingGPT: Solving {AI} Tasks with ChatGPT and its Friends in Hugging Face},'' in \emph{Advances in Neural Information Processing Systems}, 2023.

\bibitem{linkpara-4}
P.~Dighe, Y.~Su, S.~Zheng, Y.~Liu, V.~Garg, X.~Niu, and A.~H. Tewfik, ``{Leveraging Large Language Models for Exploiting ASR Uncertainty},'' \emph{{CoRR}}, 2023.

\bibitem{linkpara-5}
R.~Ma, M.~Qian, P.~Manakul, M.~J.~F. Gales, and K.~Knill, ``{Can Generative Large Language Models Perform {ASR} Error Correction?}'' \emph{{CoRR}}, 2023.

\bibitem{decoder-only-asr}
J.~Wu, Y.~Gaur, Z.~Chen, L.~Zhou, Y.~Zhu, T.~Wang, J.~Li, S.~Liu, B.~Ren, L.~Liu, and Y.~Wu, ``{On Decoder-Only Architecture For Speech-to-Text and Large Language Model Integration},'' in \emph{{ASRU}}, 2023.

\bibitem{listen_think_anderstand}
Y.~Gong, H.~Luo, A.~H. Liu, L.~Karlinsky, and J.~R. Glass, ``{Listen, Think, and Understand},'' \emph{{CoRR}}, 2023.

\bibitem{salmonn}
C.~Tang, W.~Yu, G.~Sun, X.~Chen, T.~Tan, W.~Li, L.~Lu, Z.~Ma, and C.~Zhang, ``{SALMONN: Towards Generic Hearing Abilities for Large Language Models},'' \emph{{CoRR}}, 2023.

\bibitem{qwen-audio}
Y.~Chu, J.~Xu, X.~Zhou, Q.~Yang, S.~Zhang, Z.~Yan, C.~Zhou, and J.~Zhou, ``{Qwen-Audio: Advancing Universal Audio Understanding via Unified Large-Scale Audio-Language Models},'' \emph{{CoRR}}, 2023.

\bibitem{simple-asr-llm}
Z.~Ma, G.~Yang, Y.~Yang, Z.~Gao, J.~Wang, Z.~Du, F.~Yu, Q.~Chen, S.~Zheng, S.~Zhang, and X.~Chen, ``{An Embarrassingly Simple Approach for {LLM} with Strong {ASR} Capacity},'' \emph{{CoRR}}, 2024.

\bibitem{whisper}
A.~Radford, J.~W. Kim, T.~Xu, G.~Brockman, C.~McLeavey, and I.~Sutskever, ``{Robust Speech Recognition via Large-Scale Weak Supervision},'' in \emph{{ICML}}, 2023.

\bibitem{beats}
S.~Chen, Y.~Wu, C.~Wang, S.~Liu, D.~Tompkins, Z.~Chen, W.~Che, X.~Yu, and F.~Wei, ``{BEATs: Audio Pre-Training with Acoustic Tokenizers},'' in \emph{{ICML}}, 2023.

\bibitem{libispeech}
V.~Panayotov, G.~Chen, D.~Povey, and S.~Khudanpur, ``{Librispeech: An ASR corpus based on public domain audio books},'' in \emph{{ICASSP}}, 2015.

\bibitem{hubert}
W.~Hsu, B.~Bolte, Y.~H. Tsai, K.~Lakhotia, R.~Salakhutdinov, and A.~Mohamed, ``{HuBERT: Self-Supervised Speech Representation Learning by Masked Prediction of Hidden Units},'' \emph{{ACM}}, 2021.

\bibitem{blip}
J.~Li, D.~Li, S.~Savarese, and S.~C.~H. Hoi, ``{BLIP-2: Bootstrapping Language-Image Pre-training with Frozen Image Encoders and Large Language Models},'' in \emph{{ICML}}, 2023.

\bibitem{wenetspeech}
B.~Zhang, H.~Lv, P.~Guo, Q.~Shao, C.~Yang, L.~Xie, X.~Xu, H.~Bu, X.~Chen, C.~Zeng, D.~Wu, and Z.~Peng, ``{WENETSPEECH: A 10000+ Hours Multi-Domain Mandarin Corpus for Speech Recognition},'' in \emph{{ICASSP}}, 2022.

\bibitem{aishell1}
H.~Bu, J.~Du, X.~Na, B.~Wu, and H.~Zheng, ``{AISHELL-1: An open-source Mandarin speech corpus and a speech recognition baseline},'' in \emph{{O-COCOSDA}}, 2017.

\bibitem{aishell2}
J.~Du, X.~Na, X.~Liu, and H.~Bu, ``{AISHELL-2: Transforming Mandarin ASR Research Into Industrial Scale},'' \emph{{CoRR}}, 2018.

\bibitem{aishell4}
Y.~Fu, L.~Cheng, S.~Lv, Y.~Jv, Y.~Kong, Z.~Chen, Y.~Hu, L.~Xie, J.~Wu, H.~Bu, X.~Xu, J.~Du, and J.~Chen, ``{AISHELL-4: An Open Source Dataset for Speech Enhancement, Separation, Recognition and Speaker Diarization in Conference Scenario},'' in \emph{Interspeech}, 2021.

\bibitem{speechio}
J.~Du, J.~Li, G.~Chen, and W.~Zhang, ``{SpeechColab Leaderboard: An Open-Source Platform for Automatic Speech Recognition Evaluation},'' \emph{{CoRR}}, 2024.

\bibitem{transformer}
A.~Vaswani, N.~Shazeer, N.~Parmar, J.~Uszkoreit, L.~Jones, A.~N. Gomez, L.~Kaiser, and I.~Polosukhin, ``{Attention is All you Need},'' in \emph{Advances in Neural Information Processing Systems}, 2017.

\bibitem{lora}
E.~J. Hu, Y.~Shen, P.~Wallis, Z.~Allen{-}Zhu, Y.~Li, S.~Wang, L.~Wang, and W.~Chen, ``{LoRA: Low-Rank Adaptation of Large Language Models},'' in \emph{{ICLR}}, 2022.

\bibitem{adamw}
I.~Loshchilov and F.~Hutter, ``{Decoupled Weight Decay Regularization},'' in \emph{{ICLR}}, 2019.

\bibitem{grad_clip}
J.~Zhang, T.~He, S.~Sra, and A.~Jadbabaie, ``{Why Gradient Clipping Accelerates Training: {A} Theoretical Justification for Adaptivity},'' in \emph{{ICLR}}, 2020.

\bibitem{paraformer}
{Zhifu Gao and Shiliang Zhang and Ian McLoughlin and Zhijie Yan}, ``{Paraformer: Fast and Accurate Parallel Transformer for Non-autoregressive End-to-End Speech Recognition},'' in \emph{Interspeech}, 2022.

\bibitem{u2++}
D.~Wu, B.~Zhang, C.~Yang, Z.~Peng, W.~Xia, X.~Chen, and X.~Lei, ``{U2++: Unified Two-pass Bidirectional End-to-end Model for Speech Recognition},'' \emph{CoRR}, 2021.

\bibitem{wenet}
B.~Zhang, D.~Wu, Z.~Peng, X.~Song, Z.~Yao, H.~Lv, L.~Xie, C.~Yang, F.~Pan, and J.~Niu, ``{WeNet 2.0: More Productive End-to-End Speech Recognition Toolkit},'' in \emph{Interspeech}, 2022.

\end{thebibliography}

\end{document}